\documentclass{iau} 
\usepackage{graphicx}
\usepackage{subfig}

\title[Predicting the Loci of Solar Eruptions] {Predicting the Loci of Solar Eruptions}

\author[N. Gyenge \& R. Erd\'elyi] {N. Gyenge$^{1,2,3}$ \& R. Erd\'elyi$^{1,3}$}

\affiliation{$^1$Solar Physics and Space Plasmas Research Centre (SP$^{2}$RC),\\
School of Mathematics and Statistics, University of Sheffield
\\email: {\tt n.g.gyenge@sheffield.ac.uk} \\[\affilskip]
$^2$Debrecen Heliophysical Observatory (DHO), Konkoly Observatory,\\
Research Centre for Astronomy and Earth Sciences\\ 
Hungarian Academy of Sciences, Debrecen, P.O.Box 30, H-4010, Hungary\\
$^3$Dept. of Astronomy, E\"otv\"os L. University, P\'azm\'any P. s\'et\'any 1/A\\
Budapest, H-1117, Hungary}

\pubyear{2017}
\volume{335}  
\jname{Space Weather of the Heliosphere: Processes and Forecasts}
\editors{C. Foullon \& O. Malandraki}

\begin{document}

\maketitle

\begin{abstract}

The longitudinal distribution of solar active regions shows non-homogeneous spatial behaviour, which is often referred to as Active Longitude (AL). Evidence for a significant statistical relationships between the AL and the longitudinal distribution of flare and coronal mass ejections (CME) occurrences is found in Gyenge et al. 2017 (\textit{ApJ}, 838, 18). The present work forecasts the spatial position of AL, hence the most flare/CME capable active regions are also predictable. Our forecast method applies Autoregressive Integrated Moving Average model for the next 2 years time period. We estimated the dates when the solar flare/CME-capable longitudinal belts face towards Earth.


\end{abstract}

\firstsection
\section{Motivation}

The existing flare and CME forecasting tools are usually based on the local dynamics of active regions, such as the magnetic topology (e.g., \cite[Kors\'os et al. 2014]{Korsos2014}, \cite[Kors\'os \& Erd\'elyi 2016]{Korsos2016}). However, certain global phenomena are also able to provide opportunities for predicting the locus of solar eruptive events, such as the longitudinal inhomogeneous properties of active regions (e.g., \cite[Bumba et al. 1965]{Bumba1965}; \cite[Bogart 1982]{Bogart1982}; \cite[Berdyugina et al. 2006]{Berdyugina2006}; \cite[Balthasar 2007]{Balthasar2007}; \cite[Zhang et al. 2011]{Zhang2011}). In our previous study, \cite{Gyenge2017} (hereinafter GY17), we found that the most complex active regions appear near the Active Longitude (AL). We concluded that the source of the most probably flare/CME-capable active regions is at the AL. In this paper, we combine our recently developed AL tracking method by Autoregressive Integrated Moving Average (ARIMA) model for forecasting the longitudinal position of the AL numerous Carrington Rotations (CRs) in advance.  

\section{Forecasting the Active Longitude}

\begin{figure}
	\centering
		\includegraphics[width=\linewidth]{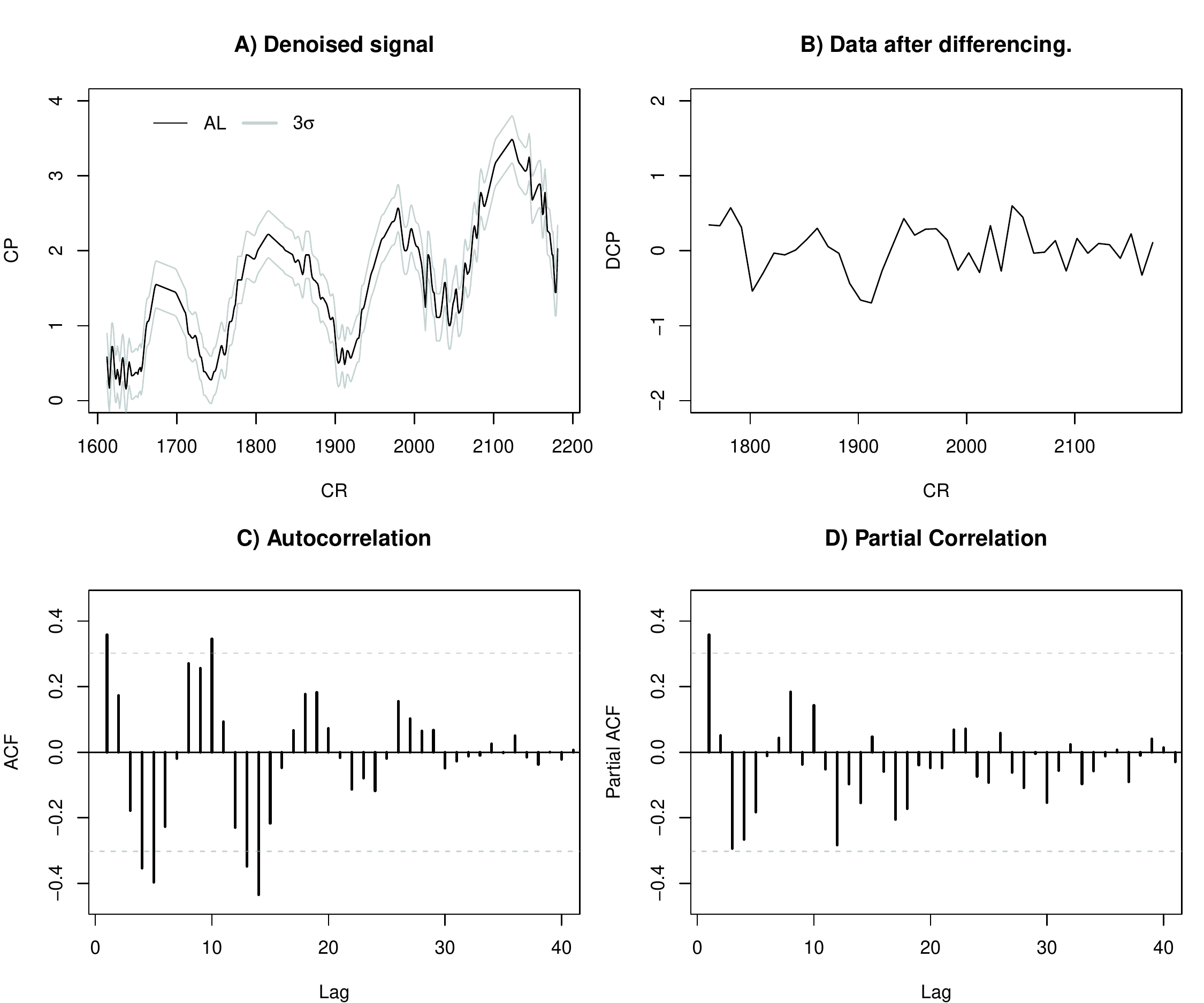}
		\caption{Migration of the AL (Panel $A$), the normalised denoised signal (Panel $B$), the autocorrelation (Panel $C$) and partial correlation (Panel $D$) of the differentiated data for the northern hemisphere are shown. One lag represents around a year.}
		\label{N_stat}
\end{figure}

We employ the Debrecen Photoheliographic Data (DPD) sunspot catalogue for determining the position of the AL. The DPD covers 540 Carrington Rotations, which equals to about five solar cycles. Panel $A$ of Figure \ref{N_stat} shows the migration of the AL. The horizontal axis represents time. The vertical axis shows the Carrington Phase ($CP = L / 360$), which is the longitudinal position $L$ of the most significant sunspot group clusters in a certain CR. The clustering method (DBSCAN) groups together points that are relatively closely packed together in a high-density region and it marks outlier points that stand alone in low-density regions (\cite[Ester et al. 1996]{Ester1996}). The longitudinal location of the clusters represent the position of the AL. The AL identification method developed based on an empirical approach, described in GY17. The data is distinguished by the hemispheres.

Seasonal ARIMA model (e.g. \cite[Ho \& Xie. 1998]{Ho1998}) is used for projecting the future values of time series, derived from the parameters ARIMA(p,d,q) $\times$ (P,D,Q). The parameters $p$ and $P$ are the number of non-seasonal and seasonal autoregressive (AR and SAR) terms. The parameters $q$ and $Q$ are the non-seasonal and seasonal moving average (MA and SMA) terms. Finally, the parameters $d$ and $D$ are the non-seasonal and seasonal differences (\cite{Box1968}). 

The application of the ARIMA model requires stationarity data; if $y_{t}$ is considered as a stationary time sample, then the distribution of any subsample $(y_{t}, ... , y_{s})$ is independent on $t$ for all $s$ (\cite{Chatfield1975}).  The signal is denoised by applying a moving average with 3 CRs window (Panel $A$ of Figure \ref{N_stat}). The black line the marks actual position of the AL clusters and the grey belt within the migration demonstrates the mean squared error. The data is non-stationary if a pronounced trend is present. Dickey-Fuller (ADF) tests are used to study the trend-stationarity. The null hypothesis of the test is that the unit root is present in a time series. The alternative hypothesis could mean trend-stationarity. The results of the test also suggest that the data is non-stationary (northern hemisphere data: \textit{p}-value $= 0.6351$ and southern hemisphere: \textit{p}-value $= 0.3054$).  

The trend-stationarity data can be archived by differencing. Now, the first difference of the time sample means the change between consecutive data points, and it is written as $y_{t}' = y_{t} - y_{t-1}$. Seasonality makes it so that the mean of the observations is not constant, but instead it evolves according to a cyclical pattern. The seasonally differenced series has a similar definition but the difference here means the difference between an observation and the corresponding observation from the previous "AL-cycle": $y_{t}' = y_{t} - y_{t-n},$ where, $n$ is the "AL-cycle". In GY17, we found that the pattern of the AL migration does not correspond well with the 11-year solar cycle. The lengths of the cycloid AL pattern iterates between 8 years and 14 years. However, there is an 11-year cyclic behaviour on average. Panel $B$ of Figure \ref{N_stat} shows the seasonal differenced and a first differenced data to obtain stationary time sample. The ADF tests confirms the results obtained with differencing and seasonal differencing (northern hemisphere: p-value $= 0.09696$ and southern hemisphere: p-value $= 0.04897$). We found that seasonal differencing $D=1$ and differencing $d=1$ are useful terms for the ARIMA model estimation.

\begin{figure}
    \centering
    \includegraphics[width=\linewidth]{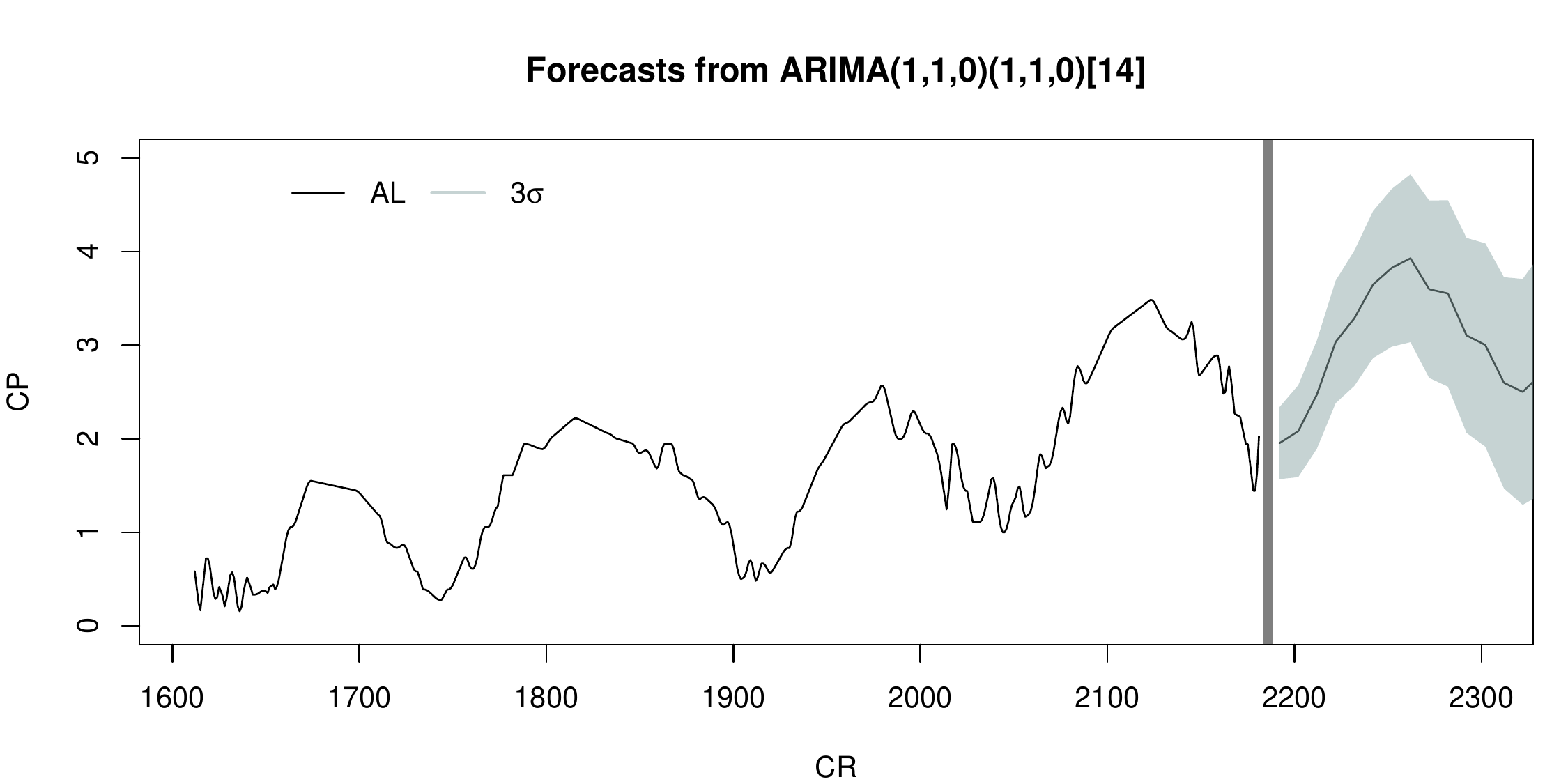}
    \caption {Forecasting the AL for the norther hemisphere. The vertical axis is the Carrington Phase ($CP$) and the horizontal axis stands for the Carrington Rotation ($CR$). The $3\sigma$ standard deviation is displayed by the grey band.}
    \label{ARIMAN}
\end{figure}

\begin{figure}
    \centering
    \includegraphics[width=\linewidth]{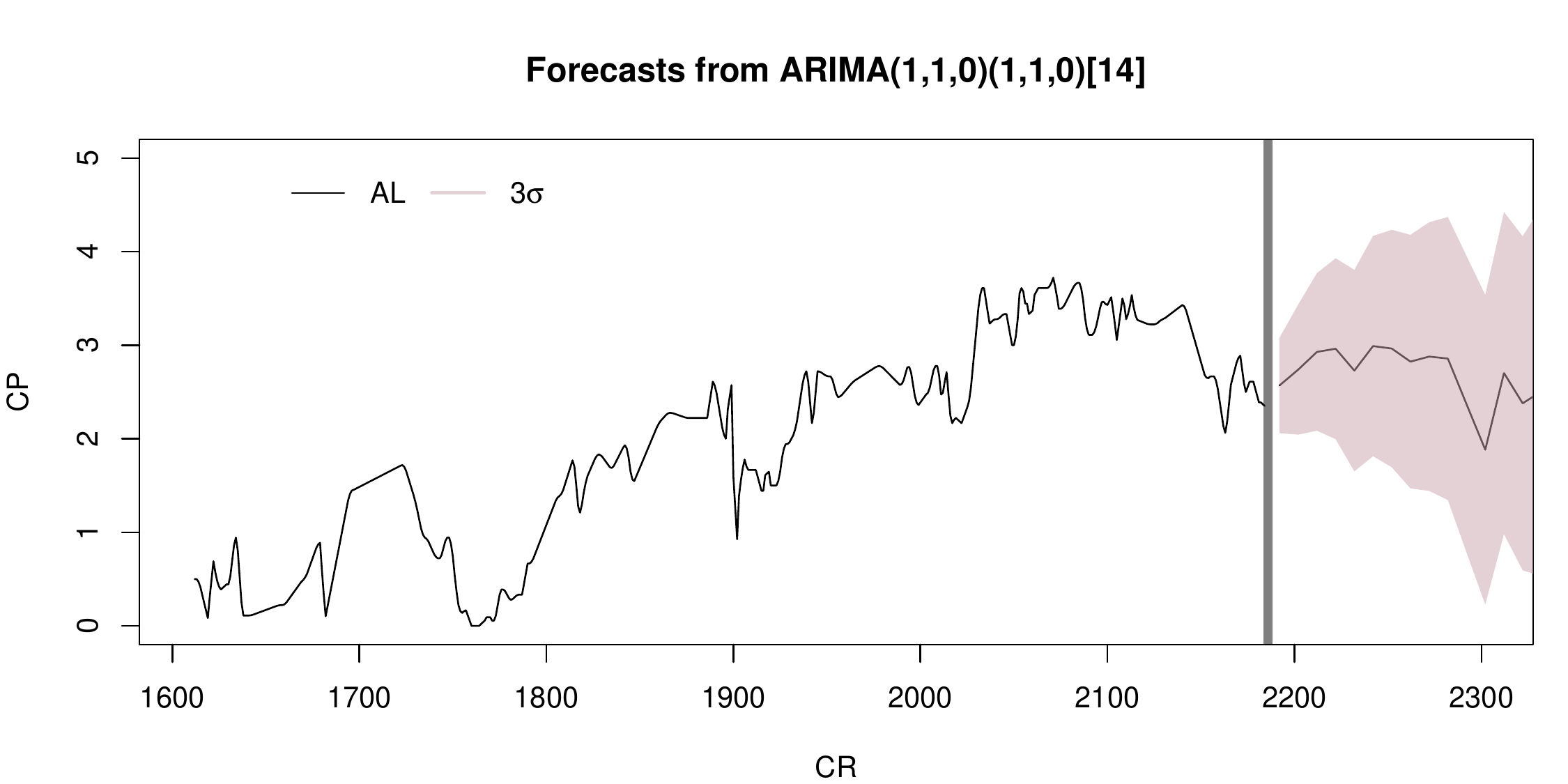}
    \caption {Forecasting the AL for the southern hemisphere. The vertical axis is the Carrington Phase ($CP$) and the horizontal axis stands for the Carrington Rotation ($CR$). The $3\sigma$ standard deviation is displayed by the red band.}
    \label{ARIMAS}
\end{figure}

The next step of the parametric estimation of the ARIMA model is the examination of the autocorrelation (ACF) and partial autocorrelation (Partial ACF) of the differenced data. Applying ACF and Partial ACF provides some initial guess value about the appropriate ARIMA model parameters (\cite[Box et al. 1994]{Box1994}).  Due the down-sampled data, lag 10 corresponds to 140 CR, which  equals to the 11-year solar cycle. Panel $C$ of Figure \ref{N_stat} (ACF) shows a harmonic wave decay pattern and the Partial ACF (Panel $D$ of Figure \ref{N_stat}) cuts off quickly. The ACF also reveals the existence of the fluctuation at positive lag around lag 10 (also 20 and 30). If the Partial ACF displays a sharp cut-off while the ACF decays more slowly, this behaviour clearly suggests an AR signature $p=1$ (\cite{Box1994}). The seasonal spikes in the ACF seem to be clear, they fade away in the ACF and cut off after lag 1 in the Partial ACF. This is the sign of a seasonal SAR model $P=1$ (\cite{Box1994}). So, it seems ARIMA(1,1,0) $\times$ (1,1,0) is a good model for estimating the future values of the time series to the northern hemisphere (Figure \ref{ARIMAN}). Similar methodology is applied for the southern hemisphere data (Panel $B$ of Figure \ref{ARIMAS}), however, the statistical error is an order of magnitude larger than in the other hemisphere. Due the large error the souther hemisphere data does not seem to provide reliable forecast.

\section{Conclusion}

By applying our method outlined in this short report, we are able to forecast the potential flare and/or CME sources several CRs in advance.  We predict that the enhanced CME and flare active longitudinal belt will face towards Earth in weeks 41, 42, 46, 50 in 2017 and weeks 2, 5, 9, 13, 17, 20, 24, 27, 35, 39 in 2018. These estimates indicate 60\% of flare and CME activity and the highest probability of fast CME occurrence (see GY17).

\section*{Acknowledgement} RE and NG are grateful to STFC and The Royal Society (UK) for the support received.

\end{document}